\documentclass[twocolumn,aps,prl,floats,showpacs]{revtex4}%
\usepackage{epsfig}
\usepackage{dcolumn}
\usepackage{bm}
\usepackage{amsmath}
\usepackage{amsfonts}
\usepackage{amssymb}
\usepackage{graphicx}%
\newcommand{\beq}{\begin{eqnarray}}
\newcommand{\eeq}{\end{eqnarray}}
\renewcommand{\vec}[1]{{\mathbf{#1}}}
\begin{document}
\title{Melting transition of an Ising glass driven by magnetic field.}
\author{L. Arrachea$^{1,*}$, D. Dalidovich$^{2}$, V. Dobrosavljevi\'c$^{2}$
and M. J. Rozenberg$^{3}$} \affiliation{$^{1}$Scuola
Internazionale Superiore di Studi Avanzati (SISSA) and Istituto
Nazionale per la Fisica della Materia (INFM) (Unit\'a di Ricerca
Trieste-SISSA), Via Beirut n. 4 I-34013 Trieste, Italy.\\
$^2$ National High Magnetic Field Laboratory,
Florida State University, Tallahassee, FL 32306, U.S.A.\\
$^{3}$Departamento de F\'{\i}sica, FCEN, Universidad de Buenos
Aires, Ciudad Universitaria Pab.I, (1428) Buenos Aires, Argentina.
}

\begin{abstract}
The quantum critical behavior of the Ising glass in a magnetic field is
investigated. We focus on the spin glass to paramagnet transition of the
transverse degrees of freedom in the presence of finite longitudinal field. We
use two complementary techniques, the Landau theory close to the $T=0$
transition and the exact diagonalization method for finite systems. This
allows us to estimate the size of the critical region and characterize various
crossover regimes. An unexpectedly small energy scale on the disordered side
of the critical line is found, and its possible relevance to experiments on
metallic glasses is briefly discussed.

\end{abstract}

\maketitle

Understanding disordered systems is one of the main challenges of condensed
matter physics, since the presence of disorder is always unavoidable in
experiments. When disorder is strong it can dominate the physics and lead to
exotic states of matter such as the glassy phases\cite{fisher}. The most
salient properties observed in glassy systems are the slow dynamical
relaxation and history dependence of thermodynamics. Research on quantum spin
systems is of primary importance because of potential technological
applications. Current work in quantum computing and spintronics,
where the understanding of relaxation processes is
crucial \cite{qcomputers,spintronics},
is boosting a renewed interest in basic models of disordered quantum
magnets.

The goal of the present work is to consider the random Ising model that
displays a quantum paramagnet to spin glass transition driven by
fluctuations introduced by an external magnetic field. We tackle the
problem by utilizing two different theoretical approaches. 
We 
solve the model using the recently introduced technique of exact
diagonalization that includes the averaging over an ensemble of disorder
realizations in a finite system. The relevant results are then obtained by
extrapolation of the data to the thermodynamic limit\cite{ar-prl,ar-prb}. This
method allows for a direct investigation of the T=0 behavior in the whole
range of parameters, circumventing thus the usual technical difficulties
encountered in the replica formalism. On the other hand, to investigate in
detail the critical behavior\cite{sachdev}, we formulate the Landau
theory in the vicinity of the quantum phase transition\cite{sach1,denvlad}.
The consistency check of results obtained using those two approaches allows us
not only to confirm their reliability, but also to identify an unexpectedly
narrow subregime near the phase boundary, in which the rapid onset of the
glassy ordering occurs. We  discuss the significance of our
findings for the current experiments on metallic glasses\cite{metalicglass}.

We consider the random Ising model that is placed in a magnetic field that
has both the transverse and longitudinal components,
\begin{align}
\label{hamil}H= 
-\sum_{<ij>}J_{ij} {\rm S}_{i}^{z} {\rm S}_{j}^{z} - \sum_{i} {\vec{h}
\cdot\vec{S}_{i}}.
\end{align}
The random interactions $J_{ij}$ are chosen to be infinite range and Gaussian
distributed with variance $J$, that sets the unit of energy in the model,
while $\vec{h}=(h^{T},0,h^{L})$. This model has an experimental realization in
the LiY$_{1-x}$Ho$_{x}$F$_{4}$ compound that has been the subject of recent
experiments \cite{aeppli1,aeppli2}. In this insulating compound, the
ground state of the magnetically active Ho ions is the low energy Ising
doublet. In addition to that, the long-range nature of dipolar interactions
between the spins enables us to perform the treatment in the large
coordination limit. Disorder in the system arises from the fact that the
substitutions of the Y atoms by the Ho ions are positionally random. The
strong randomness leads to the clear observation of the spin-glass and
ferro-glass phases at low concentration $x$ \cite{holmio}.

To investigate the transition in the system described by the Hamiltonian
(\ref{hamil}), we employ two methods that complement each other in their
scope and range of applicability. The main theoretical tool we use to obtain
the detailed analytic behavior is based on the Landau theory
approach\cite{sach1,denvlad}. Though attractive, this method is rigorously
valid only close to the quantum critical point, so that the actual range of
applicability of this approach is always difficult to assess. Hence, in
addition, we also use the exact diagonalization scheme, in which one has to
obtain the solution of $H$ for a number of explicit
realizations of disorder (typically several tens of thousands). The procedure
is implemented on finite systems of up to 17 spins. The physical
observables, such as gaps or spectral functions, are obtained along the lines
of Ref.\cite{ar-prb}. In this approach, no \textit{a priori} assumptions are
made, and its validity is limited by the reliability of the required
extrapolations to the large size limit. The main reason for success of the
previous applications of the method is that for high connectivity models the
numerical extrapolation to the thermodynamic limit is rather well behaved.
Nevertheless, as we shall see and discuss later on, in the present study we
find a certain range of parameters,where the previous statement does not hold.
Remarkably, this circumstance allows us to gain new insight into the problem.

It is useful to characterize the parameter space by $h^{L}$ and $h^{T}$, the
longitudinal and transverse components of external magnetic field $\vec{h}$
respectively. The pure transverse field case was the subject of previous
investigations \cite{ar-prl}. At T=0, the existence of the quantum phase
transition was established for a value of $h^{T} \sim O(J)$. At this point the
spin-spin dynamical local suceptibility becomes
gapless\cite{millerhuse,daniel}. When the longitudinal field is turned on, the
net longitudinal magnetization is immediately generated and the critical point
extends into a quantum critical line $h^{T}_{c}(h^{L}_{c})$. This line
separates the two phases, in which the transverse degrees of freedom of spins
are either disordered (large $h^{T}$ and $h^{L}$) or spin-glass ordered (small
$h^{T}$ and $h^{L}$). As we shall show, the excitation gap closes at this
critical line, becoming very small in some crossover region on the disordered
side of the line.

The Landau functional is constructed using the cumulant expansion about the
quantum critical point at zero longitudinal field. Both the term with random
interactions and the part with longitudinal field in the Hamiltonian 
(\ref{hamil}) are treated as perturbations. This procedure implies that the
longitudinal magnetic field $h^{L}$ is small compared to the primary
microscopic energy scale $h^{T}\sim J$. The derivation is straightforward and
leads to the following Ginzburg-Landau action\cite{sach1}
\begin{align}
& \beta\mathcal{F}=\sum_{a,\omega_{n}}\left(  \frac{r+\omega_{n}^{2}}{\kappa
}\right)  Q^{aa}(\omega_{n})+\frac{u}{2\beta}\sum_{a}\left[  \sum_{\omega_{n}%
}Q^{aa}(\omega_{n})\right]  ^{2}\nonumber\label{action}\\
& -\frac{\kappa}{3}\sum_{abc}\sum_{\omega_{n}}Q^{ab}(\omega_{n})Q^{bc}%
(\omega_{n})Q^{ca}(\omega_{n})-\frac{\beta h^{2}}{2}Q^{ab}(\omega
_{n}=0)\nonumber\\
& -\frac{\beta y}{6}\int\int d\tau_{1}d\tau_{2}\sum_{ab}\left[  Q^{ab}%
(\tau_{1}-\tau_{2})\right]  ^{4}.
\end{align}
Here $r$, being some function of $h^{T}/J$, is the parameter that governs the
transition, while $u$ and $y$ are taken at the critical point $(h^{T}%
/J)_{\mathrm{c}}\sim O(1)$. It is important to retain the quartic term,
responsible for the RSB instability\cite{denvlad}. We must insert then the
mean field ansatz
\[
\kappa Q^{ab}(\omega_{n})=\left\{
\begin{array}
[c]{ll}%
\displaystyle D(\omega_{n})+\beta q_{\mathrm{EA}}\delta_{\omega_{n},0}, &
\quad a=b,\\
\displaystyle\beta q_{ab}\delta_{\omega_{n},0}, & \quad a\neq b.
\end{array}
\right.
\]
into Eq.(\ref{action}) and vary subsequently the free energy with respect to
$D(\omega_{n})$, $q_{EA}$ and $q_{ab}$. The parametrization of $q_{ab}$
depends, however, on the phase under consideration. In the disordered
paramagnetic phase (PM) we must use the replica-symmetric ansatz
$q_{ab}=q_{EA}$, while in the spin glass phase (SG) the solution with a broken
symmetry should be used\cite{denvlad,georges}. The variational procedure is
lengthy albeit identical to that performed in the previous works. As a result,
we obtain that the equation determining $D(\omega_{n})$ is the same in both PM
and SG phases and reads
\begin{align}
& r+\omega_{n}^{2}+u\left[  \frac{1}{\beta}\sum_{\omega_{n}}D(\omega
_{n})+q_{\mathrm{EA}}\right]  -D^{2}(\omega_{n})\nonumber\label{gap}\\
& -\frac{2y}{\kappa^{2}}q_{\mathrm{EA}}^{2}D(-\omega_{n})-\frac{2y}{\kappa
^{2}}\frac{q_{\mathrm{EA}}}{\beta}\sum_{\omega_{1}}D(\omega_{1})D(-\omega
_{1}-\omega_{n})\nonumber\\
& -\frac{2y}{3\kappa^{2}}\frac{1}{\beta^{2}}\sum_{\omega_{1},\omega_{2}%
}D(\omega_{1})D(\omega_{2})D(-\omega_{1}-\omega_{2}-\omega_{n})=0.
\end{align}
This equation must be supplemented by
\beq\label{edwan}
2D(0)q_{\mathrm{EA}}+\frac{2y}{3\kappa^{2}}q_{\mathrm{EA}}^{3}+\frac
{h^{2}\kappa}{2}=0
\eeq
in the PM phase and
\beq\label{qglass}
q_{\mathrm{EA}}^{2}=-[D(0)\kappa^{2}]/y
\eeq
in the SG phase, to comprise the full system to be solved self-consistently.
Though the exact treatment of this system is not possible, we can obtain the
leading order of the correct solution close to the quantum critical point. We
consider here only the case of $T=0$, so that all the sums over Matsubara
frequencies are substituted by the corresponding integrals.

We notice first that, if $y=0$, the complete solution is easily derived to be
\cite{sach1} $D(\omega_{n})=-\sqrt{\omega_{n}^{2}+\Delta^{2}}$. The gap
$\Delta^{2}$, that turns to zero right at the critical point, is determined
using the following identity
\begin{align}
\label{iden}\int\frac{d\omega}{2\pi} (\omega^{2} +\Delta^{2})^{1/2}=
\frac{\Lambda_{\omega}^{2}}{2\pi}+\frac{\Delta^{2}}{2\pi} \ln(c_{1}
\Lambda_{\omega}/\Delta)
\end{align}
In Eq.(\ref{iden}) $\Lambda_{\omega}$ is the upper frequency cutoff and
$c_{1}$ is some constant of order unity. Let's assume that for $y\ne0$ the
leading approximation of $D(\omega_{n})$ contains the same square root
singularity as for $y=0$, and analyze how the last two terms in Eq.(\ref{gap})
affect the solution in the leading approximation. Simple inspection reveals
that in the prelast term it is sufficient to put $\omega_{n}=0$, $\Delta=0$
while calculating the integral over $\omega_{1}$. This contributes only to the
renormalization of the coefficient $u$ before $q_{\mathrm{EA}}$, so that
$uq_{\mathrm{EA}}\rightarrow u_{1} q_{\mathrm{EA}}$.

The last term requires, however, the calculation of the integral
\begin{align}
& K(\Delta,\omega_{n})=\int\frac{d\omega_{1}}{2\pi}\int\frac{d\omega_{2}}%
{2\pi}\sqrt{\omega_{1}^{2}+\Delta^{2}}\sqrt{\omega_{2}^{2}+\Delta^{2}%
}\nonumber\label{K}\\
& \times\sqrt{(\omega_{1}+\omega_{2}+\omega_{n})^{2}+\Delta^{2}},
\end{align}
that is difficult to perform exactly for arbitrary $\omega_{n}$ and
$\Delta^{2}$. We need, however, only the leading behavior of this integral
provided $\omega_{n},\Delta\ll1$. A simple estimate yields:
\beq\label{smallK}
K(\Delta,\omega_{n})=A+B\omega_{n}^{2}+C_{1}\Delta^{2}\ln(C_{2}/\Delta),
\eeq
where the constants $A,B,C_{1}$ and $C_{2}$ are some cut-off 
$\Lambda_{\omega}$ dependent functions. We see that the first term in 
the above expression
renormalizes the critical value $r_{c}$ (equal to $u\Lambda_{\omega}^{2}/2\pi$
for $y=0$), while the contribution from the second one can be simply absorbed
by the appropriate rescaling of temperature $T$ in $\omega_{n}^{2}$. The third
term in Eq.(\ref{smallK}) leads to the renormalization of the coefficient
before the $\Delta$-dependent part of Eq.(\ref{iden}).

\begin{figure}[ptb]
\epsfxsize=3.5in  \epsffile{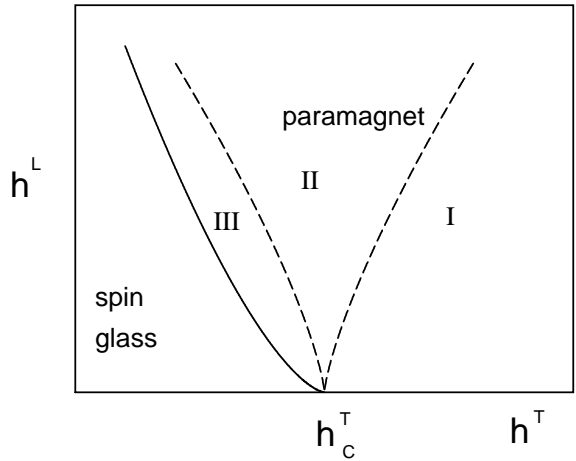}
\caption{Schematic phase diagram predicted by the Landau theory. The dashed
lines denote crossovers while the full line is a critical line.}%
\label{fig1}%
\end{figure}

Similarly as in Ref.\cite{denvlad}, we obtain that in the PM phase
\begin{align}
\label{D}D(\omega_{n})= -y q_{\mathrm{EA}}^{2}/\kappa^{2}- \sqrt{\omega
_{n}^{2}+\Delta^{2}},\nonumber\\
\Delta^{2}=\displaystyle \frac{ r-r_{c} +u_{1} q_{\mathrm{EA}} } { u_{2} \ln[
C u_{2} / ( r-r_{c} +u_{1} q_{\mathrm{EA}} ) ] },
\end{align}
where $C, u_{1}$ and $u_{2}$ are again some $\Lambda_{\omega}$ dependent
functions of the order unity. As a result of solution of Eqs.(\ref{gap}) and
(\ref{edwan}), one can distinguish the following regimes on a ($r-r_{c}$, $h$)
plane (see Fig.\ref{fig1}). 

(I) In this regime, in which $h \ll(r-r_{c})^{3/4}$, $q_{\mathrm{EA}}$ is the
smallest parameter and can be treated as a perturbation. As a result, we
obtain with the logarithmic accuracy, that $q_{\mathrm{EA}}=(\kappa h^{2}
)/4\Delta$, $\Delta\approx\{(r-r_{c}) / u_{2} \ln(1/(r-r_{c}) \}^{1/2}$. This
equation shows that when $h^{L}$ becomes non zero, $q_{\mathrm{EA}}$ also
becomes finite even in the PM phase due to the finite magnetization along the
longitudinal axis.

\begin{figure}[ptb]
\epsfxsize=3.in  \epsffile{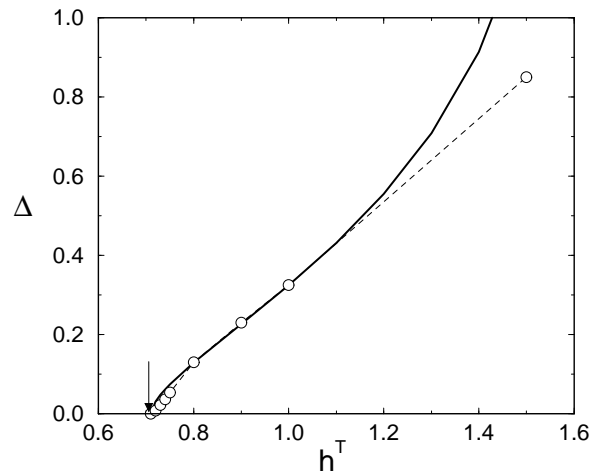}
\caption{Gap vs. transverse field $h^{T}$ at $h^{L}=0$ (open circles). The
fitting function from Eq.(\ref{D}) with $q_{\mathrm{EA}}$=0 is plotted in the
solid line. The arrow indicates the critical field.}%
\label{fig2}%
\end{figure}

The expression for the gap was first obtained in Ref.\cite{millerhuse} and
\cite{sach2} that considered the $h^{L}$=0 case. To answer the question of the
region of validity of the Landau approach, we use the exact diagonalization
method to obtain the gap as a function of $h^{T}$ at $h^{L}$=0. The results
are shown in Fig.\ref{fig2}. The agreement at small values of $\Delta$
demonstrates the reliability of our methods and gives an indication of the
size of the critical region.

(II) This region is characterized by the condition $|r-r_{c}|^{3/4} \ll h$. In
the leading approximation $\Delta\approx\{ (u_{1} \kappa h^{2}) / 4u_{2}
\ln(1/h^{4/3}) \}^{1/3}$, while $q_{\mathrm{EA}} \approx\{ (\kappa h^{2} /4)
\sqrt{ (u_{2} /u_{1}) \ln(1/h^{4/3}) } \}^{2/3}$.

(III) This regime, in which $(r_{c}-r)^{3/4} \gg h$, is the closest to the
$T=0$ critical boundary. The EA order parameter, 
that crosses over to its value in
the glassy phase, is given by $q_{\mathrm{EA}}=[(r_{c} -r)/u_{1}]+(u_{2}
\Delta^{2} /u_{1}) \ln\left[ 1/\Delta^{2}\right]  $, with  $\Delta
\approx[\kappa u_{1} h^{2}/4(r_{c}-r)]- [2y(r_{c} -r)^{2}/ 3 u_{1}^{2}
\kappa^{2}]$.  From this expression it is easily seen that $\Delta$ vanishes
at the critical line given by $h=(8y/3)[(r_{c}-r)/u_{1} \kappa]^{3/2}$.

Finally, in the SG phase:
\begin{align}
\label{dglass}D(\omega_{n})= -y q_{\mathrm{EA}}^{2} /\kappa^{2} -|\omega
_{n}|,\quad q_{\mathrm{EA}}=(r_{c}-r)/u_{1},
\end{align}
resulting in a gapless form of the spectral density $\mathrm{Im}\chi
(\omega)\propto\omega$.

\begin{figure}[ptb]
\epsfxsize=3.in  \epsffile{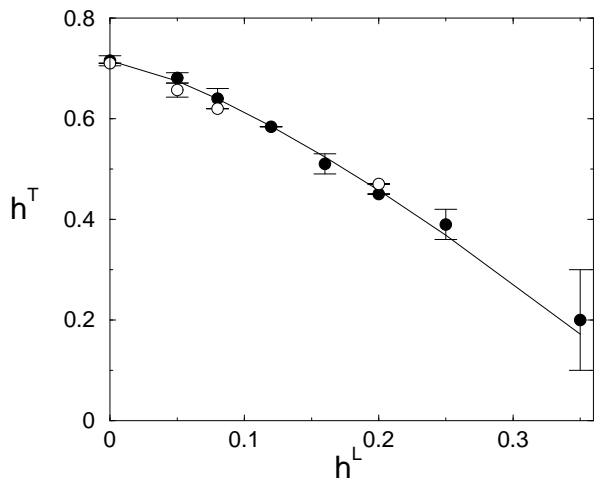}
\caption{SG-PM phase boundary obtained with exact diagonalization. Filled and
open circles correspond to the two different criteria (i) and (ii),
respectively (see text). The solid line corresponds to the fitting function
$h^{T}=h_{c}^{T}-2.2(h^{L})^{3/4}$.}%
\label{fig3}%
\end{figure}

We would like now to discuss the nature of the crossover between subregimes II
and III in more detail. A rather surprising result, one obtains from the exact
diagonalization method, is that in fact the freezing transition of the
transverse degrees of freedom takes place at the critical boundary line given
by $h_{c\mathrm{ED}}^{T}\propto|h^{L}-h_{c}^{L}|^{3/4}$ (see Fig. \ref{fig3}).
This result was verified by two different criteria: (i) the divergence of the
spin-glass susceptibility given by $J^{2}\langle\lbrack\chi_{loc}^{zz}%
]^{2}\rangle=1$, and (ii) the vanishing of the excitation energy gap of the
regular part of the dynamical spin susceptibility, that corresponds to the so
called "replicon" mode\cite{georges}. It is notable that the infinite
system size extrapolations for these two different freezing transition criteria
do agree well. However, these results seem paradoxical since the Landau theory
predicts a phase transition boundary with a different functional form, namely,
$h_{c}^{T}\propto|h^{L}-h_{c}^{L}|^{3/2}$ (and different curvature, see Fig.
\ref{fig2}). 

This paradox is resolved upon further scrutiny of the results from the
Landau theory. To this end, it is important to note that, in the presence of
the non-zero longitudinal field, the critical behavior of the gap is different
than at $h^{L}=0$. It takes a much slower, linear form $\Delta(\delta
r)\sim\delta r$ ($\delta r$ is the distance to the critical line), becoming
the new effective small energy scale that characterizes the region III. This
linear regime of $\Delta(\delta r)$ crosses over to the regime II, at values
of $r_{c}-r \approx( \kappa u_{1} h^{2}/4)^{2/3} u_{2}^{1/3} \ln^{1/3}
(1/h^{4/3})$. Remarkably, this is precisely the functional form obtained for
the critical line (and gap closure) from the numerical calculation.

The key point is that for systems of the size that one can diagonalize, the
physics of the small gap is masked by the finite size effects, affecting thus
the validity of extrapolations. This is a well known limitation of exact
diagonalization studies that occurs in systems in which the small energy
scales emerge\cite{hm-1d}. Interestingly, similar discrepancies are found in
the prediction of the de Almeida - Thouless (AT) line in the classical
Sherrington-Kirkpatrik (SK) model (with $h^{T}=0$) in a longitudinal field.
Numerical calculations suggest that the critical temperature is
 $T_{c} \propto|h^{L}-h^{L}_{c}|^{3/4}$, instead
of the correct result $T_{c} \propto|h^{L}-h^{L}_{c}|^{3/2}$. In the classical
case, this can be an indication that the free energies of the ordered and
paramagnetic phases are actually very close within a crossover region in the
$T-h^{L}$ diagram, equivalent to the region III of Fig. 1. The large
finite-size effects were found also in the recent numerical simulations of the
classical model \cite{skfiniteT}, and are possibly relevant to the anomalous
behavior observed in experimental studies of the AT line \cite{mal}. For the
quantum systems, such small gaps may be also difficult to observe in experiments
as well as in numerical calculations. In contrast, in regions I and
II, the $r$-dependence of $\Delta$ assumes a form similar to the zero field
limit (see Eq.(\ref{D}), except that $r$ is shifted by the
quantity $u_{1} q_{\mathrm{EA}}$). Since in region II (dropping logarithmic
corrections) $q_{\mathrm{EA}}\sim h^{4/3}$, we conclude that the crossover
line separating regions II and III may play a role of an \emph{apparent}
critical line, below which the gap, although finite, may assume unobservable
small values.

This outstanding feature, which was overlooked in previous works, may have
important consequences. For instance, it may be responsible for the peculiar
observation of the quenching of the nonlinear susceptibility at the quantum
critical point of the LiY$_{1-x}$Ho$_{x}$F$_{4}$ series \cite{wu}. Another
example is the electron glass model that was recently described in Ref.
\cite{denvlad} and for which essentially identical arguments apply. In this
case the dynamical exponent is $z=1$, and we find that the crossover energy
scale (corresponding to the gap in the Ising case) behaves as $\Delta
\sim\delta r^{2}$ and corresponds to a crossover temperature separating the
Fermi liquid regime (at low $T$) from the quantum critical regime (at high
$T$). The second power in $\delta r$ indicates an even broader quantum
critical regime than in the Ising case. Such an extended quantum critical
region may result in enhanced dissipation at low temperatures, a possibility
which may bear relevance for the puzzling absence of weak localization
(interference) corrections in certain two-dimensional electron gases in the
low density regime.

We thank S. Sachdev for useful discussions and suggestions. 
This work was supported by the
National High Magnetic Field Laboratory (DD and VD), NSF grants
DMR-9974311 and DMR-0234215 (VD), CONICET (LA and MJR) UBACyT and Fudaci\'on
Antorchas (MJR).


\begin{thebibliography}{999}                                                                                              %
\bibitem[(*)]{address} 
$^*$Permanent address: Departamento de F\'{\i}sica, 
Universidad de Buenos Aires, Ciudad Universitaria Pabell\'on I, (1428) Buenos
Aires, Argentina.

\bibitem {fisher}K.H.Fischer and J.A.Hertz, Spin Glasses, Cambridge University
Press, Cambridge, England (1991).

\bibitem{qcomputers}G.Santoro, {\em et al},
Science, \textbf{295}, 2427-2430 (2002)

\bibitem {spintronics}M.L.Roukes, Nature \textbf{411}, 747-748 (2001).

\bibitem {ar-prl}L.Arrachea and M.J.Rozenberg, Phys. Rev. Lett. \textbf{86},
5172 (2001).

\bibitem {ar-prb}L.Arrachea and M.J.Rozenberg, Phys. Rev. B \textbf{65},
224430 (2002).

\bibitem {sachdev}S. Sachdev. Quantum Phase Transitions, Cambridge University
Press, Cambridge, England (1999).

\bibitem {sach1}N. Read, S. Sachdev, J. Ye, Phys. Rev. B \textbf{52}, 384 (1995).

\bibitem {denvlad}Denis Dalidovich and Vladimir Dobrosavljevi\'{c}, Phys. Rev.
B, (Rapid Comm.) \textbf{66}, 081107(2002).

\bibitem {metalicglass}S. Bogdanovich and D. Popovi\'{c}, Phys. Rev. Lett.
\textbf{88}, 236401 (2002); Phys. Rev. Lett. \textbf{89}, 289904 (2002); J.
Jaroszynski, D. Popovi\'{c}, and T.M. Klapwijk, Phys. Rev. Lett. \textbf{89},
276401 (2002).

\bibitem {aeppli1}J.Brooke, {\em et al}, Science
\textbf{284}, 779 (1999).

\bibitem {aeppli2}J.Brooke, T.F.Rosenbaum and G.Aeppli,
Nature \textbf{413}, 610 (2001).

\bibitem {holmio}D.H.Reich \textit{et al.}, Phys. Rev. B \textbf{42}, 4631
(1990). W. Wu \textit{et al.}, Phys. Rev. Lett. \textbf{67}, 2076 (1991).

\bibitem {millerhuse}J.Miller and D.A.Huse, Phys. Rev. Lett. \textbf{70}, 3147 (1993).

\bibitem {daniel}M.J.Rozenberg and D.R.Grempel, Phys. Rev. Lett. \textbf{81},
2550 (1998).

\bibitem {georges}A. Georges, O. Parcollet, S. Sachdev, Phys. Rev. B
\textbf{63}, 134406 (2001).

\bibitem {sach2}S. Sachdev and J. Ye, Phys. Rev. Lett. \textbf{70}, 3339 (1993).

\bibitem {hm-1d}C.A.Stafford, A.J.Millis and B.S.Shastry, Phys. Rev. B \textbf{43},
13660 (1991); C.A.Stafford and A.J.Millis,Phys. Rev. B \textbf{48}, 1409 (1993); 
L. Arrachea, E. R.
Gagliano and A. A. Aligia, Physica C \textbf{268} 233 (1996); Phys. Rev. B
\textbf{55}, 1173 (1997).

\bibitem {skfiniteT}A. Billoire and B. Coluzzi, cond-mat/0302026.

\bibitem {mal}A. P. Malozemoff, S. E. Barnes and B. Barbara, Phys. Rev. Lett.
\textbf{51} 1704 (1983).

\bibitem {wu}W. Wu, {\em et al} Phys. Rev. Lett.
\textbf{71}, 1919 (1993); J. Mattsson, Phys. Rev. Lett. \textbf{75}, 1678
(1995); D. Bitko, T. F. Rosenbaum, and G. Aeppli, Phys. Rev. Lett.
\textbf{75}, 1679 (1995).
\end{thebibliography}
\end{document}